\documentstyle[11pt,newpasp,twoside,epsf]{article}

\begin{document}

\title{Magnetic Fields in Young Galaxies}
\author{{\AA}ke Nordlund and \"{O}rn\'{o}lfur R\"{o}gnvaldsson}
\affil{Theoretical Astrophysics Center and Nordita, Copenhagen}

\begin{abstract}
We have studied the fate of initial magnetic fields in the hot halo gas
out of which the visible parts of galaxies form, using three-dimensional
numerical MHD-experiments.  The halo gas undergoes
compression by several orders of magnitude in the subsonic cooling
flow that forms the cold disk. The magnetic field is carried along and
is amplified considerably in the process, reaching $\mu$G levels
for reasonable values of the initial ratio of magnetic to thermal energy
density.
\end{abstract}

% ---------------------- intro ---------------------------------------
\section{Introduction}

The origin of large scale magnetic fields in disk galaxies is an
unsettled issue. Proposed solutions to this problem fall in two categories.
On the one hand, in-situ dynamo action has been invoked to
exponentially amplify an initially weak, small scale, seed field
(for recent reviews, see Beck et al. 1996; Kulsrud 1999).
\nocite{Beck+Brandenburg+Moss+;1996,Kulsrud;1999}
On the other hand, large scale pregalactic
fields in the halos of forming galaxies might be compressed and stretched
sufficiently during the formation and evolution of galactic disks to
explain the observed large scale fields (see
e.g.~Howard \& Kulsrud 1997).
\nocite{Howard+Kulsrud;1997}

Observations of $\mu$G level large scale fields in slowly
rotating irregular galaxies (Chy{\.z}y et al.~2000),
\nocite{Chyzy+Beck+Kohle+;2000}
high redshift objects (e.g.~Perry et al.~1993),
\nocite{Perry+Watson+Kronberg+;1993}
and in the intracluster medium (Eilek 1999; Colgate \& Li 2000),
\nocite{Eilek;1999,Colgate+Li;2000}
pose fundamental problems for the classical
$\alpha-\Omega$ galactic dynamo as a cornerstone for
creating magnetic fields on galactic (and larger) scales.
In addition, it seems difficult to attribute the strong vertical field
found in the central molecular zone of our Galaxy to
a galactic dynamo, which may be
suggestive of a primordial origin for this component of the galactic
magnetic field (Sofue \& Fujimoto 1987; Chandran, Cowley, \& Morris 2000).
\nocite{Sofue+Fujimoto;1987,Chandran+Cowley+Morris;2000}

The observational evidence for magnetic fields
in cluster halos and the association between these fields and outflows
from AGNs suggests that similar processes may seed protogalactic halos
with large scale magnetic fields.
The compression and stretching of the field during the galaxy
formation process further amplifies the halo field and could possibly
account for the observed fields in present day disk galaxies.
The conventional criticism that this kind of
scenarios will result in an incorrect parity with respect to the
galactic plane may be shown to result from overly simplified
assumptions about the initial field.   Only in the case of
a field that is both weakly inclined and initially located
near the center of the potential well does it follow that the
wound-up field is predominantly of odd parity.  %Such initial
%conditions are not to be expected if the large scale halo
%fields originate from contamination by AGN jets.
%%AA  This sentence cut only to squeeze down to three pages.

\begin{figure}
\plottwo{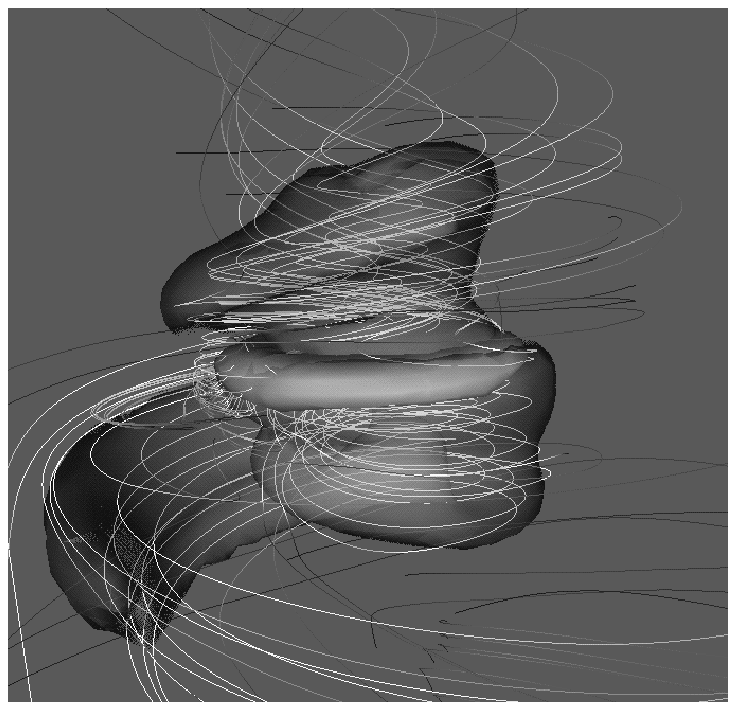}{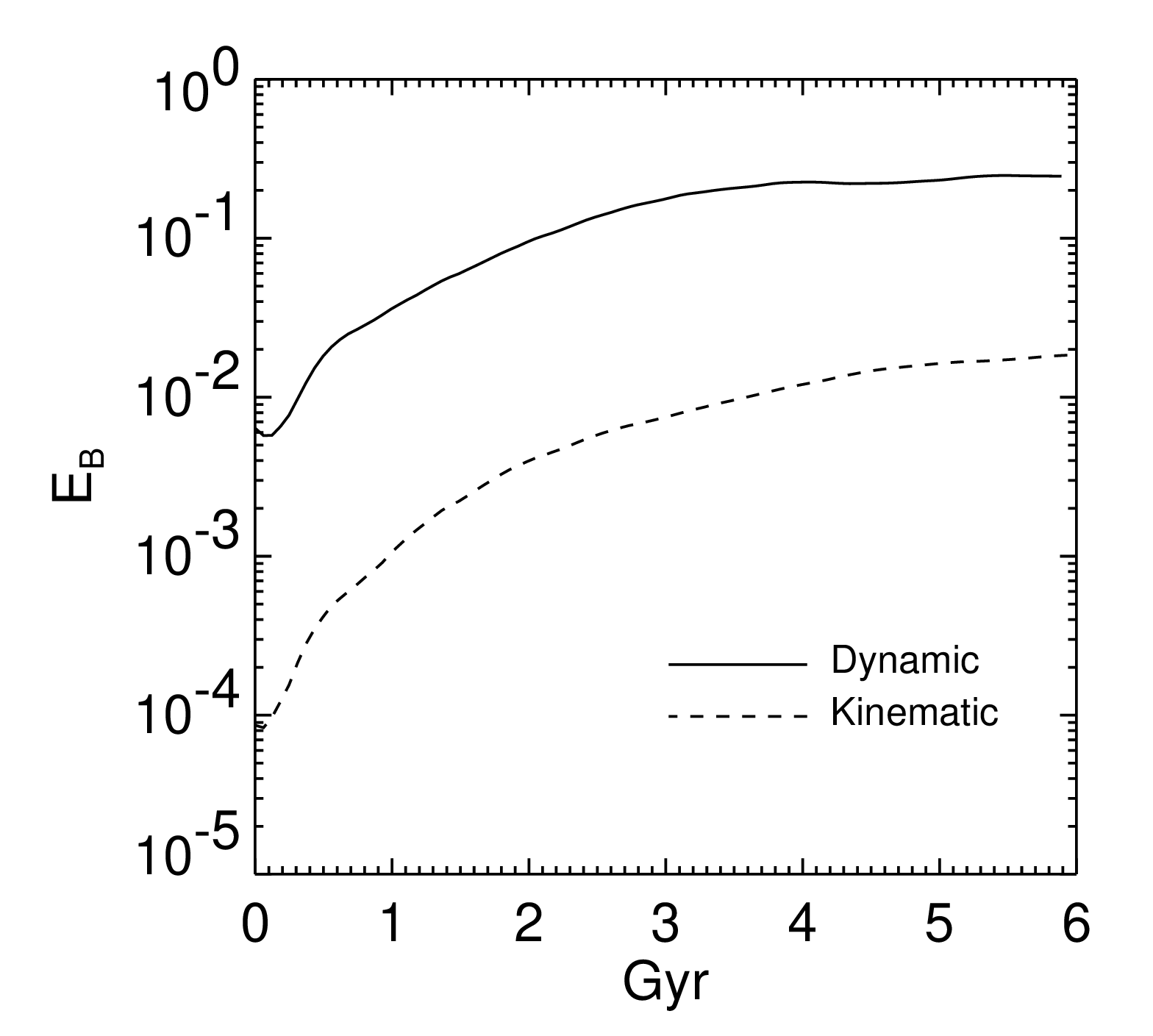}
\caption[]{
(a) The magnetic energy isosurface (dark
gray) shows the asymmetry of the field as it gets
dragged into the cold disk (light gray) with the cooling flow
(for 3-D color renderings see
\verb'www.astro.ku.dk/'${\mathtt\sim}$\verb'aake/talks/IAU2000').
Field lines show that the polarity of the field changes in
the halo, above the plane of the disk.
(b) The growth of the average magnetic energy in a volume enclosing
the cold disk and the halo gas immediately above and below the
disk. The initial magnetic pressure in the ``dynamic'' experiment is
1\% of the thermal pressure, while the kinematic case starts off
with 0.01\%.
}
\label{fig:Btopo}
\end{figure}

This has prompted us to study the compression and wind-up of initial
halo fields during the build-up of a disk galaxy in a cooling flow.
By adopting a smooth background potential and by using sufficient
numerical resolution (enhanced by grid stretching) we achieve that
the disk forming at the
bottom of the potential well has the characteristic size, mass, and
specific angular momentum of a typical disk galaxy (the initial spin
parameter $\lambda = {J E^{1/2} / G M^{5/2}}$ is equal to 0.056).
The hot halo gas is threaded
by a random magnetic field with approximately constant $P_{\rm
mag}/P_{\rm gas}$, plus a large scale component of similar energy
content that has a non-zero inclination and offset from the rotation
axis.

\section{Results and Discussion}
\label{sec:discussion}

If the initial field has a significant
inclination and offset from the center, the resulting wound-up
field has a reversal that is offset into the halo (see Fig.~1a).
When observed
face-on, in optically thin radio frequencies, the magnetic field
would appear to be uni-directional, as is normally observed,
because one of the winding directions dominates over the other.
The reversal in the halo could be revealed by analyzing observations in different
radio frequencies, as has indeed been done for M51 (Berkhuijsen et al.~1997).
\nocite{Berkhuijsen+Horellou+Krause+;1997}
Note that, in the model presented here, there is no creation
of ``new'' magnetic flux---the growth of magnetic energy (Fig.~1b) is
purely a result of compression and winding of an
(approximately conserved) amount of magnetic flux, initially
threading the disk plane over a range of radii.  With sufficient
numerical resolution of the disk dynamics, there would of course be
additional fine scale structure and enhancement of the magnetic energy,
associated with turbulence in the disk.

The other common point of criticism against galactic magnetic
fields originating from large scale, pre-galactic fields is the
issue of pitch angle.  Here it should be remarked that conventional
mean field dynamos invoke a large amount of magnetic diffusion,
particularly in spiral arms,
together with a rapid regeneration of the radial magnetic
field component, to maintain the pitch angle.  The magnetic diffusion
is assumed to arise, for example, from turbulence driven by star
formation.  If such a mechanism was allowed for in the current
experiments, it would be equally effective in maintaining the
pitch angle.  Regeneration of the radial field component is not
necessary; apart from vertical transport effects, its magnitude
is determined by the (conserved) amount of wound-up magnetic flux
and the pitch angle.
It has also been argued that galactic winds, caused by the same SNe that
drive the turbulence, could be a significant source of loss of magnetic
field from galactic disks.  However, as has been demonstrated by
Gudiksen (1999),
\nocite{Gudiksen;1999}
the hot wind emanates
from the disc in ``chimneys'', which are intermixed with much denser
regions that are able to hold on to the magnetic field.
Loss of magnetic
flux is also counteracted by the incoming cooling flow.

In summary, the scenario adopted here is consistent both with
current ideas about early AGNs as the origin of the observed,
large scale magnetic fields in clusters, and with observed
properties of galactic magnetic fields.  Since the results
were obtained by simply tracing the magnetic field evolution in
a cooling flow that results in a disk galaxy which is
in other respect consistent with observations, one may
even turn the argument around and conclude that the initial
$P_{\rm mag}/P_{\rm gas}$ of the halo gas would have to be much
smaller than the $P_{\rm mag}/P_{\rm gas}$ typical of large scale
cluster magnetic fields, in order not to give rise to galactic
magnetic fields similar to the observed ones.

% ------------------------ References ----------------------------

%\references

\end{document}